\documentclass[twocolumn,showpacs,preprintnumbers,amsmath,amssymb,prl,superscriptaddress]{revtex4-2}
\usepackage{multirow}
\usepackage{amsfonts}
\usepackage[english]{babel}
\usepackage[T1]{fontenc}
\usepackage{times}
\usepackage{mathrsfs}
\usepackage{graphicx}
\usepackage{dcolumn}
\usepackage{bm}
\usepackage[colorlinks,bookmarks=true,citecolor=blue,linkcolor=red,urlcolor=blue]{hyperref}
\usepackage[tight, FIGTOPCAP, hang, raggedright, nooneline]{subfigure}

\begin{document}

\title{Zero-Clustering Geometry in Realistic Fractional Quantum Hall Wave Functions}

\author{Xin Wan}
\email{xinwan@zju.edu.cn}
\affiliation{Zhejiang Institute of Modern Physics and 
Zhejiang Key Laboratory of Micro-Nano Quantum Chips and Quantum Control, 
Zhejiang University, Hangzhou 310027, China}

\author{Ziang Wang}
\affiliation{Zhejiang Institute of Modern Physics and 
Zhejiang Key Laboratory of Micro-Nano Quantum Chips and Quantum Control, 
Zhejiang University, Hangzhou 310027, China}

\author{Zi-Xiang Hu}
\email{zxhu@cqu.edu.cn}
\affiliation{Department of Physics and Chongqing Key Laboratory for Strongly Coupled Physics,
Chongqing University, Chongqing 401331, China}

\author{Zhao Liu}
\email{zhaol@zju.edu.cn}
\affiliation{Zhejiang Institute of Modern Physics and 
Zhejiang Key Laboratory of Micro-Nano Quantum Chips and Quantum Control, 
Zhejiang University, Hangzhou 310027, China}

\date{\today}

\begin{abstract}
The clustering pattern of zeros in the ground state of a fractional quantum Hall system is a defining feature of its topological properties.
We analyze the geometrical fluctuations of the zeros around individual electrons and propose to use the displacement ratio of the zeros 
to visualize and measure the distance of a realistic state to a model wave function. The distribution of the zero displacement ratio behaves like an order parameter in the transition from a Laughlin phase to a topologically trivial one. 
The statistical comparison between quantum Hall states belonging to different Jain sequences leads to a composite fermion fluid description of the $\nu = 1/5$ ground state with long-range Coulomb interaction that agrees almost perfectly for as few as 
$3$-$5$ electrons, overcoming the long-standing difficulties of accommodating the competing liquid and crystal orders at short distances.
\end{abstract}

\maketitle

{\it Introduction.} Quantum phases of strongly interacting electrons host correlations far more complex than those required by Fermi statistics alone. Shortly after the discovery of the celebrated fractional quantum Hall (FQH) effect, Laughlin proposed a many-body wave function with a plasma analogy 
to capture the correlations among electrons~\cite{laughlin83}. 
Haldane pointed out that the Laughlin wave function is the exact ground state of a projective pseudopotential Hamiltonian~\cite{haldane83},
while Halperin emphasized that its zeros play a defining role of the electron correlation 
in the sense that if we freeze the positions of all but one electron at filling fraction $\nu = 1/m$, it sees $m$ zeros at the positions of 
the frozen ones~\cite{halperin83}. Only one of them is required by the Pauli principle, whereas the other $m-1$ 
non-Pauli zeros reflect the strong correlation that keeps electrons apart. The binding of zeros to the frozen electrons echoes the idea of 
composite fermions (CFs)~\cite{jain89}, in which each electron binds an even number of quantized vortices and 
senses a smaller effective magnetic field. 

Distinct clustering structure of the zeros~\cite{moore91,read99} can be introduced by  generalized projection 
Hamiltonians~\cite{greiter92,simon07}. Up to a ubiquitous Gaussian factor, the resulting ground-state wave functions are 
holomorphic functions of the complex electron coordinates in the lowest Landau level (LLL), 
which can, alternatively, be constructed by conformal field theory~\cite{moore91} or the Jack polynomial approach~\cite{bernevig08a,bernevig08b}.
Wen and Wang emphasized that the topological properties of Abelian and non-Abelian
FQH states can be classified by the pattern of zeros extracted from symmetric polynomials of infinite variables~\cite{wen08a,wen08b}.
Thus, unlike in the Landau symmetry-breaking description of phase transitions, 
the breakdown of the long-range entanglement in FQH states may be tied to the evolution of the zero clustering structure.

Such evolution of the zero structure and the subsequent phase transition have been revealed quantitatively in anisotropic FQH systems~\cite{haldane11,qiu12}. The model anisotropic wave functions, geometrically deformed by mass and/or interaction 
anisotropy, describe the continuous evolution of the zero structure characterized by 
metric parameters~\cite{qiu12}. For the family of anisotropic $\nu = 1/3$ Laughlin states, the degenerate triple zeros 
for each electron split; the non-Pauli zeros depart along a direction fixed by anisotropy, leading to a stretched correlation hole.
The relative ratio of the displacement of the non-Pauli zeros to the average distance among electrons 
can be used to locate the transition from the anisotropic Laughlin fluid to a gapless liquid-crystal phase~\cite{wang12}.

Beyond the model FQH states, Yoshioka observed nondegenerate zeros in Coulomb ground states with 
Halperin and Lee~\cite{yoshioka83}, while he emphasized in his book that it is the strong Coulomb interaction at short distance 
that leads to the binding of zeros to electrons~\cite{yoshiokabook}.
The residual effect of the ubiquitous Coulomb interaction forces the zeros to repel each other in the ground state that is different from but still close to the Laughlin state. 
Laughlin argued that the slight difference can be attributed to short-range forces between zeros in the equivalent plasma picture, 
which have no effect on the neutrality of the plasma~\cite{laughlin_in_book}.

However, unlike in the case of model FQH states where the existence of analytical wave functions makes the zero structures clear, the distribution of zeros in realistic wave functions remains elusive, and several key questions naturally arise.
How can one quantify the deviation from the ideal zero structure of the model Laughlin state, especially when the perturbation is not small?
Can such a measure diagnose the topological order of the Laughlin state, 
in particular, through a phase transition into a topologically trivial state? More ambitiously, can the geometrical fluctuations of zeros guide us to construct better variational 
wave functions for realistic systems?

In this Letter, we answer these questions by analyzing the statistics of the non-Pauli zeros clustered around individual electrons in FQH systems. 
We propose that the zero displacement ratio, which compares complex 
electron-zero spatial displacements, is a sensitive probe to measure
the proximity between a realistic state and FQH model wave functions. 
We analyze its density evolution during a breakdown of the Laughlin order, establishing its effectiveness in detecting topological phase transitions. 
Furthermore, we use the density evolution to analyze a realistic $\nu = 1/5$ system with Coulomb interaction, 
in which case the Laughlin topological order competes with the Wigner crystal (WC) order, which tends to arrange electrons in a lattice to maximally reduce long-range Coulomb interaction.
We show that a CF fluid wave function emerging from such an analysis describes the realistic state consistently better than that involves 
artificial crystallite embedding even for systems of 3-5 electrons. 

\begin{figure}
\begin{center}
\includegraphics[width=8cm]{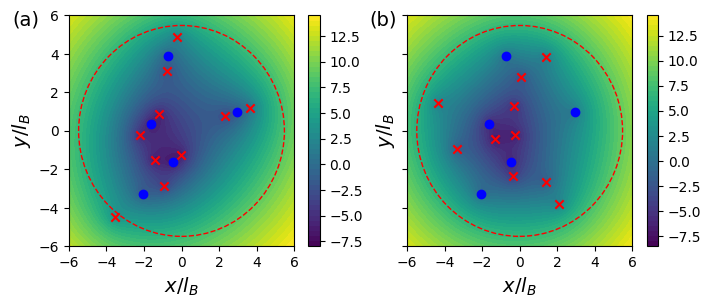}
\end{center}
\caption{
\label{fig:zeros_dist}
(a) Location of the ground-state wave function zeros of an $N = 6$ and $\nu = 1/3$ system 
with Coulomb interaction. The positions of five electrons are fixed (blue dots), 
around which the additional 10 zeros (red crosses) appear in pairs. 
The background is the filled contour plot of the logarithm of 
the wave function amplitude with the five electron position fixed. 
(b) Similar plot for the ground state of Eq.~(\ref{eq:hamiltonian}) with 
$\lambda = 1.4$. 
}
\end{figure}

{\it Statistics of zeros in the $\nu=1/3$ Laughlin phase.} We first consider the FQH system with $N$ electrons at $\nu=1/3$ in disk geometry. 
The electrons are interacting via a combination of the short-range 
and Coulomb interaction 
\begin{equation} 
    H(\lambda) = (1 - \lambda) H_0 + \lambda H_C, 
\label{eq:hamiltonian}
\end{equation}    
where $H_0$ contains the pseudopotential $V_1 = 1$ only  
and $H_C$ is the Coulomb interaction projected onto the LLL.  
We solve the ground state wave function in the Laughlin momentum sector via exact diagonalization (ED). The $\lambda=0$ limit gives the model $\nu=1/3$ Laughlin state.
We then sample $N - 1$ electron positions from a Poisson distribution 
within a disk of radius $R = \sqrt{2N/\nu}$~\cite{notes_on_sampling} and solve the wave-function zeros for the remaining electron. 

An example of the collection of fixed electrons and non-Pauli zeros 
for the six-electron Coulomb ground state is plotted in a complex plane 
in Fig.~\ref{fig:zeros_dist}(a).
For comparison, we plot the non-Pauli zeros for its ground state of 
the mixed Hamiltonian with $\lambda = 1.4$, which is no longer in the Laughlin phase~\cite{haldane1985}, 
with the same set of fixed electron positions in Fig.~\ref{fig:zeros_dist}(b).
The stark contrast confirms that the distribution of zeros, in particular its nearly linear triple-zero structure,
is the hallmark of the Laughlin topological order. 
To quantify such a pattern in $\nu = 1/3$ Laughlin-like states, we define the zero displacement ratio for the two nearest neighboring (NN) zeros 
around an electron as
\begin{equation}
\label{eq:zeta}
\zeta = \frac{z_1 - z_e}{z_2 - z_e},
\end{equation}
where the complex $z_e$ denotes the location of the electron, 
and $z_1$ and $z_2$ its NN and the next NN zeros, respectively. 
The definition is partly inspired by the introduction of the real ratio of 
the consecutive level spacings in disordered systems of interacting electrons~\cite{oganesyan07}. 
The complex ratio $\zeta$ has the following properties:
(1) $\zeta$ is bound by the unit circle, and $\vert \zeta \vert = 1$ 
indicates the equidistance of the two nearest zeros to the electron.  
(2) Avoidance of non-electron zeros from each other is reflected by $\zeta \neq 1$.
(3) The tendency of two zeros sitting symmetrically opposite to each other around an electron 
is parametrized by $\zeta = -1$, which is the feature of the Laughlin-like states, 
regardless of the intrinsic geometry of the wave functions~\cite{haldane11,qiu12}. 

\begin{figure}
\begin{center}
\vspace{0.5cm}
\includegraphics[width=8cm]{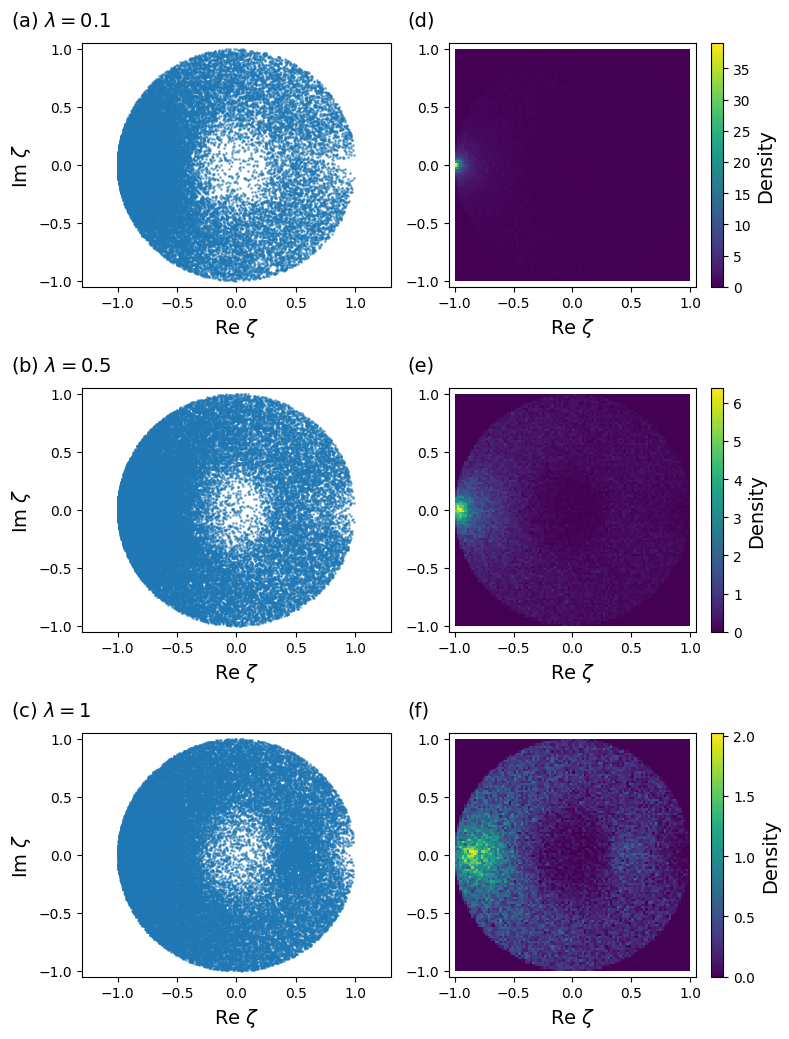}
\vspace{-0.5cm}
\end{center}
\caption{
\label{fig:zero_displacement_ratio}
The scatter plot of $\zeta$ in 10,000 realizations of randomly fixed electron positions
in the $N = 6$ and $\nu = 1/3$ system. 
The mixture of $V_1$ and Coulomb interactions is (a) $\lambda = 0.1$, (b) $\lambda = 0.5$, 
and (c) $\lambda = 1$, and the corresponding density maps are shown in (d)-(f), respectively. 
}
\end{figure}

In Fig.~\ref{fig:zero_displacement_ratio} we plot $\zeta$ for 10,000 realizations 
of zeros and their density for the ground states of an $N = 6$ electron system 
with various mixed $V_1$ and Coulomb interaction. 
In all cases, the ratio tends to avoid $\zeta = 1$ due to the repulsion among the zeros.
It also tends to avoid $\zeta = 0$, which is due to the repulsion between zeros 
and electrons when the ground state departs from the exact Laughlin form. 
When the interaction is dominated by the short-range $V_1$ potential, e.g., at $\lambda = 0.1$, 
the density of $\zeta$ sharply peaks at $\zeta \approx -1$, 
consistent with the nearly symmetric drift of two non-Pauli zeros around an individual electron along diametrically opposite directions caused by a small perturbation from the pseudopotential~\cite{kusmierz21}.
As $\lambda$ increases, the peak gradually broadens and shifts away from $-1$ along the real axis.
The shift of the peak location quantifies the asymmetry between the two 
nearest zeros around the same electron, which comes from the repulsion from the zeros in the neighboring electron-zero clusters~\cite{sm}. 
The repulsion, in general, drives the zeros away from their corresponding electrons 
along different directions, resulting in complex values of $\zeta$. However, the peak is still located near $\zeta = -1$ even in the pure Coulomb case, signaling the persistence of the Laughlin order.

{\it Evolution of zeros during the collapse of the $\nu=1/3$ Laughlin phase.} While the sharp peak near $\zeta = -1$ is the hallmark of the Laughlin phase, 
its flattening and disappearance can be associated with the destruction of the Laughlin topological order.
For a demonstration, we consider a simpler mixed interaction $H(\eta)=(1-\eta)V_1+\eta V_3$. 
At $\eta=0$, the ground state is the model $\nu=1/3$ Laughlin state. 
The Laughlin order survives at small $\eta$ when $V_1$ dominates over $V_3$. 
Nevertheless, it is replaced by a charge order (stripe, smectic, or bubble phase) at sufficiently large 
$\eta$~\cite{haldane1985,nicolas_nematic,bubble_phase}.

Here we switch to the torus geometry to facilitate the comparison with 
conventional phase-transition diagnostics like the energy gap, 
and to avoid the edge effects. The positions of $N-1$ electrons are uniformly sampled on the torus. The magnetic translation symmetry guarantees 
an exact three-fold degeneracy of the ground states, regardless of the underlying many-body phase. 
Given the coordinates of $N-1$ electrons, we can numerically identify three zeros that shift with the ground-state momentum, 
whereas all other zeros remain identical across the ground-state manifold. 
These two classes of zeros can be naturally attributed to
the contributions from the center of mass and the relative coordinate of electrons
in the many-body wavefunction, respectively.
Such a factorization of the many-body wavefunction has been explicitly demonstrated for the Laughlin model state~\cite{haldane_torus,ronny_torus}.
We therefore calculate $\zeta$ from the zeros shared by the degenerate ground states.
The shortest distance on the torus is used to measure the separation 
between an electron and a zero. 
 
\begin{figure}
\centerline{\includegraphics[width=\linewidth]{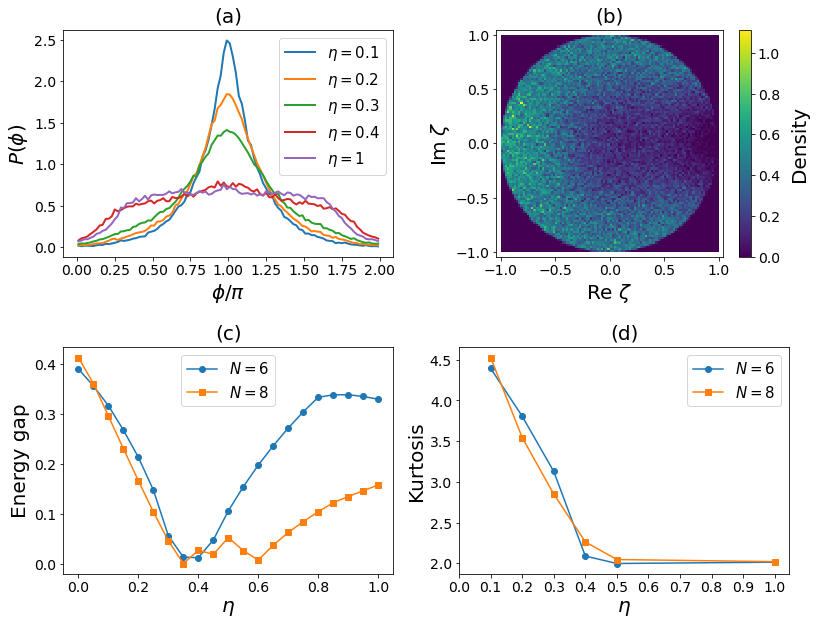}}
\caption{(a) The distribution of $\zeta$'s argument $\phi$ at various $\eta$. 
(b) The density map of $\zeta$ at $\eta=0.4$. (c) The ground-state energy gap versus $\eta$.  
(d) The kurtosis of $P(\phi)$ versus $\eta$. The system size is $N=6$ electrons in (a) and (b). }
\label{fig:v1v3}
\end{figure}

At small $\eta$, we indeed observe the Laughlin-type clustering pattern of an electron 
and its two nearest non-Pauli zeros, as found on the disk geometry. 
The peak near $\zeta=-1$ can be characterized by 
the distribution $P(\phi)$ of $\zeta$'s argument $\phi\in[0,2\pi)$. 
As shown in Fig.~\ref{fig:v1v3}(a), $P(\phi)$ has a sharp peak at $\phi=\pi$ at small $\eta$,
signaling the persistence of the Laughlin phase. 
Notably, $P(\phi)$ transits to a different shape at $\eta_c\approx 0.4$. 
When $\eta\geq \eta_c$, the pronounced peak in $P(\phi)$ is replaced by a wide
plateau, meaning that the two nearest zeros no longer tend to sit opposite to the electron
as in the Laughlin state, although they still avoid 
collinearity with the electron on the same side. Consistently, the distribution of $\zeta$ is significantly broadened [Fig.~\ref{fig:v1v3}(b)], 
indicating a much stronger asymmetry between the two nearest zeros around the electron.
These striking changes in the zero configuration point to the collapse of the $\nu=1/3$ Laughlin topological order. 
Remarkably, the ground-state energy gap vanishes at almost the same $\eta_c$ [Fig.~\ref{fig:v1v3}(c)], which confirms that the phase transition 
from the $\nu=1/3$ Laughlin state is indeed quantitatively detected by the configuration pattern of zeros.
We further quantify the evolution of $P(\phi)$ by its kurtosis $K$.
With increasing $\eta$, $K$ drops to about $2$ at $\eta\approx 0.4$ and then saturates [Fig.~\ref{fig:v1v3}(d)], 
behaving like an order parameter to characterize the phase transition.
The saturation value is close to $K=1.8$ that corresponds to a uniform distribution of $\phi$,
indicating the collapse of Laughlin-type electron-zero clustering. 

{\it Composite-fermion fluids at $\nu=1/5$.} The success in applying the zero displacement ratio to the identification 
of the $\nu = 1/3$ Laughlin phase and its transition to a topologically trivial phase
motivates us to explore FQH states with more complex zero structures. To be concrete, we consider the $\nu=1/5$ filling, for which we still adopt the Hamiltonian with the form of Eq.~(\ref{eq:hamiltonian}) but include two pseudopotentials $V_1 = V_3 = 1$ in the $H_0$ term. The ground state in the $\lambda=0$ limit is the model $\nu=1/5$ Laughlin state. The sampling method of $N-1$ electrons is the same as that at $\nu=1/3$.

It is well known that the $\nu = 1/5$ ground state for pure Coulomb interaction 
is an incompressible fluid state, but its proximity to the WC phase 
makes it substantially different from the model Laughlin state~\cite{lam84,yang01,lauchli10}.
This proximity is reflected in the possibility of forming 
so-called type-2 CF crystals on the background of an FQH fluid away from the $1/5$ filling~\cite{archer13}. 
Exactly at $\nu = 1/5$, strong short-range crystalline correlations 
are present in the ground state, making the CF crystal wave function constructed by 
a Hartree-Fock crystal of electrons bound with two flux quanta~\cite{PhysRevB.58.4019,chang05,chang06}
surprisingly superior to the Laughlin liquid wave function for $N < 10$ 
electrons~\cite{chang06}. 

\begin{figure}
\centering
\includegraphics[width=\linewidth]{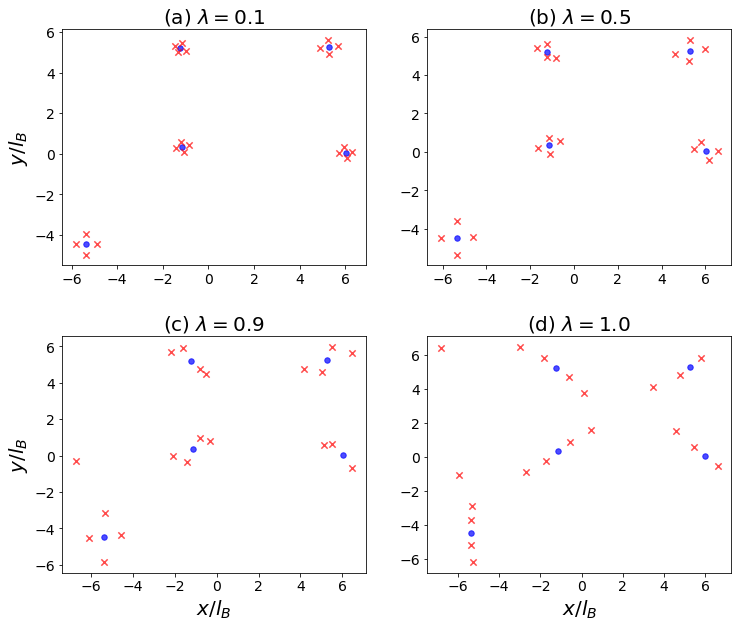}
\caption{The evolution of electron positions (blue dots) and non-Pauli zeros (red crosses) 
with the increasing of $\lambda$ at $\nu=1/5$ on the torus for a specific configuration of electrons.
The system size is $N=6$ electrons. In (d),
electrons which are expected at one side of the sample may appear on the opposite side 
due to the periodic boundary conditions on the torus.
}
\label{fig:one_fifth_torus}
\end{figure}

The model Laughlin wave function at $\nu = 1/5$ has four zeros bound to each electron. 
As exemplified in Fig.~\ref{fig:one_fifth_torus}(a) for almost no Coulomb interaction 
($\lambda = 0.1$) on the torus, these zeros tend to form two equidistant pairs along perpendicular directions.
Nevertheless, with the increasing of $\lambda$ to the pure Coulomb value $\lambda=1$, 
the star-like pattern of zeros undergoes a crossover to a one-dimensional crystal-like structure, 
as shown in Figs.~\ref{fig:one_fifth_torus}(b)-(d).
During this evolution, while the nearest two non-Pauli zeros remain tightly 
bound to the electron, 
the other two appear to meander around farther away
and align with the closer zeros with the pure Coulomb interaction.
Such a crossover is also featured in the collective evolution 
of a generalization of the zero displacement ratio~\cite{sm}.
The pattern evolution strongly suggests that the usual picture of the Laughlin phase 
at $\nu = 1/5$ that each electron evenly binds four zeros cannot be stabilized 
by the Coulomb interaction.
Its long-range part tends to arrange the zeros unequally;
only two zeros or vortices are tightly bound to each electron,
highlighted in Fig.~\ref{fig:one_fifth_torus}(d),  
forming $^2{\rm CF}$s at an effective filling fraction $\nu^* = 1/3$.  
Subsequently, these $^2{\rm CF}$s form a hierarchical $\nu^*=1/3$ CF fluid (CFF) state, 
whose loosely bound zeros effectively screen the long-range interaction
among electrons so other zeros can be tightly bound.
This is a manifestation of geometrical fluctuations in the FQH effect~\cite{haldane11}
in the absence of an extrinsic symmetry breaking field. 

\begin{figure}
\centering
\includegraphics[width=\linewidth]{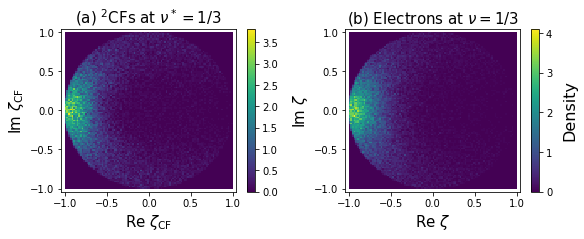}
\caption{The density maps of $\zeta$ for (a) $^2{\rm CF}$s at $\nu=1/5$ ($\nu^*=1/3$)
and (b) electrons at $\nu=1/3$.
The system size is $N=6$ electrons on the torus.
}
\label{fig:CF_one_third_torus}
\end{figure}

The hierarchy picture guides us to compare the zero clustering statistics of 
the $\nu = 1/3$ Coulomb ground state and the $\nu^* = 1/3$ CFF state. For the latter,
the center-of-mass coordinate of an electron 
and its two NN zeros determines the position of a CF, and the two next NN zeros of the electron are used as $z_{1,2}$ in Eq.~(\ref{eq:zeta}) for the statistical analysis. 
The remarkable similarity in the density maps of $\zeta$ 
(see Fig.~\ref{fig:CF_one_third_torus})
strongly supports the formation of the $\nu^*=1/3$ fluid state of $^2{\rm CF}$s.
It suggests that an alternative CF construction of the $\nu = 1/5$ FQH state
from a fluid state can outperform that from a crystal state as in Ref.~\cite{chang06}
even in very small systems.
Hence, we define the CFF state by
\begin{equation}
    \Psi_{\rm CFF} = \prod_{i<j} (z_i - z_j)^2 \Psi_{\rm fluid}^{1/3},
\end{equation}
where $\Psi_{\rm fluid}^{1/3}$ is the ground state 
of Eq.~(\ref{eq:hamiltonian}) at $\nu^* = 1/3$ with a variational parameter $\lambda$. 

In Table~\ref{tab:wf_comparison}, we compare the overlap amplitudes 
$\mathcal{O}$ between the CFF states constructed at different $\lambda$'s 
and the Coulomb ground state at $\nu = 1/5$ on the disk. 
The $\nu=1/5$ model Laughlin state can also be regarded as a CFF state constructed 
with $\lambda=0$. 
Here, we apply ED without any orbital cutoff, 
followed by exact algebraic flux attachment, for up to $N = 6$ electrons. 
Table~\ref{tab:wf_comparison} confirms that the Coulomb CFF state constructed 
with $\lambda=1$ outperforms the model Laughlin wave function. 
The variation of $\lambda$ allows us to further improve the overlap between the CFF state and the Coulomb ground state.
The optimal tuning $\lambda_{\rm opt} =  1.091 \pm 0.008$ is almost size-independent and close to the critical point~\cite{sm,haldane1985} to a neighboring charge-ordered phase at $\nu^*=1/3$~\cite{haldane1985}, but the $\nu^*=1/3$ parent state is still a fluid. Heuristically, at $\lambda=\lambda_{\rm opt}$ one can think that the short-range part of the residual interaction 
between $^2{\rm CF}$s is suppressed from the electron case, due to the screening 
by the two vortices tightly bound to each electron. The proximity to the critical point introduces long-range crystalline correlations that the Coulomb interaction may cause to complement the liquid phase, thus improving the overlap with the Coulomb ground state. Compared with the artificial mixture of 
the CF crystal state and the Laughlin state~\cite{chang06} which introduces strong crystalline correlations 
to the 1/5 state in an alternative approach, our CFF wave function at $\lambda_{\rm opt}$ shows consistently better overlaps for $N = 3$-$5$. In principle, our $N = 6$ result can be further improved by tuning more
short-range pseudopotential components.
Therefore, our CFF construction reveals the true 
thermodynamic nature at tiny sizes. 

\begin{table}
\caption{The overlap amplitude $\mathcal{O}$ of the model Laughlin state, the Coulomb CFF (CCFF), 
and the variational CFF (VCFF) with the exact Coulomb ground state at $\nu = 1/5$
for up to $N=6$ electrons on the disk. The VCFF state is constructed from 
the $\nu^*=1/3$ parent ground state with $\lambda = \lambda_{\rm opt}$
that maximizes the overlap with the exact ground state.}
\begin{center}
\setlength{\tabcolsep}{10pt} 
\def\arraystretch{1.2}
\begin{tabular}{ccccc}
\hline \hline
$N$ & $\mathcal{O}_{\rm Laughlin}$ & $\mathcal{O}_{\rm CCFF}$ & 
$\lambda_{\rm opt}$ &$\mathcal{O}_{\rm VCFF}$ \\
\hline
3 & 0.985392 & 0.995733 & 1.088 & 0.999988 \\
4 & 0.947491 & 0.983465 & 1.085 & 0.999131 \\
5 & 0.909861 & 0.954991 & 1.083 & 0.997742 \\
6 & 0.842390 & 0.893517 & 1.099 & 0.985935 \\
\hline \hline
\end{tabular}
\end{center}
\label{tab:wf_comparison}
\end{table}

{\it Conclusions.} To summarize, we study the distribution of wave-function zeros 
in Laughlin-like states from a geometrical point of view. 
We show that the zero displacement ratio, 
informed by the level spacing ratio in disordered quantum systems, 
can be used to characterize the topological order of the Laughlin phase
and to detect its transition to other phases.
The correlation of the zeros provides us with
new insights into the CF construction of the FQH states and 
their geometrical description beyond static deformation 
due to global or local anisotropy. 
In particular, there is a hierarchical structure of zeros 
in the family of generic $\nu = 1/5$ FQH states,
among which the Laughlin wave function can be regarded as 
the consequence of zero coalescence in the absence of 
realistic interaction and disorder. 

Therefore, zero-clustering geometry is an indispensable tool in 
studying the long-range entanglement and phase transitions in realistic systems.
We envisage that the statistical tool can shed new light on the 
FQH systems with anisotropic mass or interaction, disordered potential, Landau-level mixing, and other realistic perturbations.
Another interesting investigation is to uncover how non-Abelian statistics 
manifests in the distribution of zeros in realistic FQH systems. 
Extending this zero-clustering framework to fractional Chern insulators (FCIs)~\cite{Sun-PhysRevLett.106.236803,Tang-PhysRevLett.106.236802,neupert-PhysRevLett.106.236804,sheng-natcommun.2.389,regnault-PhysRevX.1.021014,Parameswaran2013816,BERGHOLTZ-JModPhysB2013,LIU2024515,xie2021fractional,xu2023fractional2,jie2023fractional,xu2023fractional,li2023fractional,ju2023fractional,li_fci_review}, the lattice analogues of FQH states at zero magnetic field,
represents a particularly exciting direction for future research. 
We anticipate that the geometrical analysis of zeros developed in this work will offer a novel 
diagnostic of topological orders in FCIs, and help elucidate their connection to conventional FQH states, 
particularly in regimes where the Chern number or quantum geometry of 
the hosting Chern bands deviates significantly from the Landau-level limit.

\section{ACKNOWLEDGMENTS}

This work was supported by the National Key Research and Development Program of China Project No.~2021YFA1401902 
(X.~W., Z.~W. and Z.~L.). Z.~L. was also supported by the  National Natural Science Foundation of China (Grant No.~12350403 and 12374149).
Z.-X.~H. was supported by the National Natural Science Foundation of China (Grant No.12474140 and No.12547101) and the Fundamental
Research Funds for the Central Universities Grant No. 2025CDJ-IAISYB-029.

\bibliography{BibFQH}

\begin{thebibliography}{50}
\bibitem [{not()}]{notes_on_sampling}%
  \BibitemOpen
  \href@noop {} {}\bibinfo {note} {Monte Carlo sampling of electron positions
  suggests that their distribution is more uniform than the uncorrelated
  Poisson distribution~\cite{laughlin_in_book}. However, the effect is weak as
  we are interested in the displacement between an electron and its nearest
  zeros.}\BibitemShut {Stop}%
\end{thebibliography}

\newpage
\renewcommand\thefigure{S\arabic{figure}}
\renewcommand\theequation{S\arabic{equation}}
\setcounter{equation}{0} 
\setcounter{figure}{0} 
\setcounter{table}{0}  

\onecolumngrid

\section{Supplementary Material for ``Zero-Clustering Geometry in Realistic Fractional Quantum Hall Wave Functions''}

In this supplementary material, we present detailed numerical data in disk geometry, including 
the zero displacement ratio of a three-electron Laughlin state at $\nu = 1/3$,
the detailed evolution of the ratio distribution at $\nu = 1/5$, and
the construction of composite fermion fluid wave function for Coulomb interaction at $\nu = 1/5$.

\subsection{I. Zero Displacement Ratio in Three-Electron Laughlin States at $\nu = 1/3$ in Disk Geometry}

\begin{figure*}[b]
\begin{center}
\includegraphics[width=16cm]{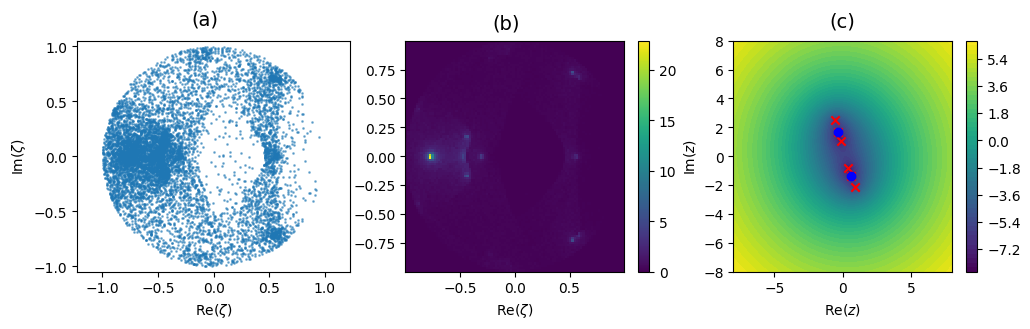}
\end{center}
\caption{
\label{smfig:n3_zero_map}
(a) Scatter plot of zeros for 10,000 
random realizations of two fixed electron positions 
in the three-electron Coulomb ground state. 
(b) The density map of zeros of 100,000 random realizations. 
The highest or brightest peak is located at $\zeta = -0.77$.
(c) A typical configuration of zeros which yield $\zeta = -0.77$.
The background is the colormap for $\ln \vert \psi (z, z_e^{(2)}, z_e^{(3)}) \vert$, 
the logarithm of the (unnormalized) wave function amplitude.
The red crosses indicate the locations of the zeros, when two electrons are fixed at the blue dots. 
}
\end{figure*}

We find in the main text that the zeros of a Laughlin-type wave function exhibit 
a dominant peak near $\zeta = -1$, and the peak location shifts along the real axis 
as $\lambda$ increases. 
We argue that such a shift reflects the interaction among electrons and 
their neighboring zeros. 
Here, we use a three-electron system in disk geometry with Coulomb interaction as an example to 
illustrate the origin of the $\zeta$ peak and the large deviations from it. 
Strictly speaking, in the case of $N = 3$, there is no separation of bulk and edge;
neither is well defined.
Nonetheless, we can view the two fixed electrons as a part of a larger system, 
in which the rest of the $N - 3$ fixed electrons are separated  
from the two-electron cluster by a distance much larger than 
the two-electron separation.
Therefore, the $\zeta$ peak of a generic $N$-electron system can be approximated,
hence understood, by that of the $N = 3$ system.
Fig.~\ref{smfig:n3_zero_map}(a) shows the scattering of zeros for 10,000 
random configurations of two fixed electron positions. 
The majority of the zeros fall to the left of the origin, 
indicating that two nearest non-Pauli zeros tend to align opposite to each other 
around each electron. 
Fig.~\ref{smfig:n3_zero_map}(b) shows the density map of zeros of 
100,000 random realizations. 
The map reveals a sharp peak at $\zeta_0 = -0.77$.
In addition, there are weaker spots scattered on both sides of the origin. 
In the following, we will explore the dependence of the features on 
fixed electron positions. 

In general, we can write the $\nu = 1/3$ wave function after we fix the positions of 
$N - 1$ electrons (which we denote as $z_e^{(2)}, z_e^{(3)} , \dots, z_e^{(N)}$) as 
\begin{equation}
\Psi(z) = \prod_{i = 1}^{3(N-1)} (z - w_i),  
\end{equation}
up to an overall factor, where $w_i$s locate the resulting $3(N-1)$ zeros, 
including the $N-1$ electrons.  
As we consider a rotationally symmetric system in disk geometry, 
the wave function is homogeneous. 
Consequently, if we scale each electron location by $\lambda$ such that 
$z_e^{(i)} \rightarrow \lambda z_e^{(i)}$, the wave function simply becomes 
\begin{equation}
\Psi_{\lambda}(z) = \prod_{i = 1}^{3(N-1)} (z - \lambda w_i),  
\end{equation}
which means that each zero is scaled by the same factor $\lambda$. 
As a result, the zero displacement ratio for the $i$th electron
\begin{equation}
\zeta^{(i)} = \frac{z^{(i)}_1 - z_e^{(i)}}{z^{(i)}_2 - z_e^{(i)}}
\end{equation}
remains invariant, where $z^{(i)}_j$ is the $j$th NN zero of the $i$th electron. 
The scale invariance simplifies our discussion on $N = 3$, 
as we only need to consider dependence of $\zeta^{(2)}$ and $\zeta^{(3)}$ 
on the magnitude and angle of $\chi = z_e^{(3)} / z_e^{(2)}$. 
In addition, we can always choose the second electron to be the closer one 
such that $\vert \chi \vert \geq 1$. 

\begin{figure*}[b]
\begin{center}
\includegraphics[width=6cm]{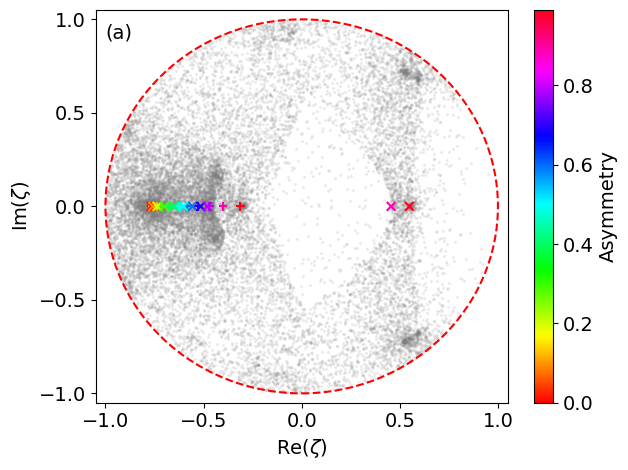}
\includegraphics[width=6cm]{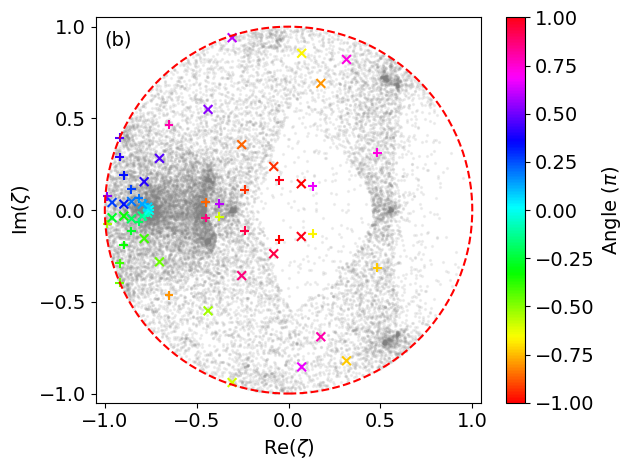}
\includegraphics[width=6cm]{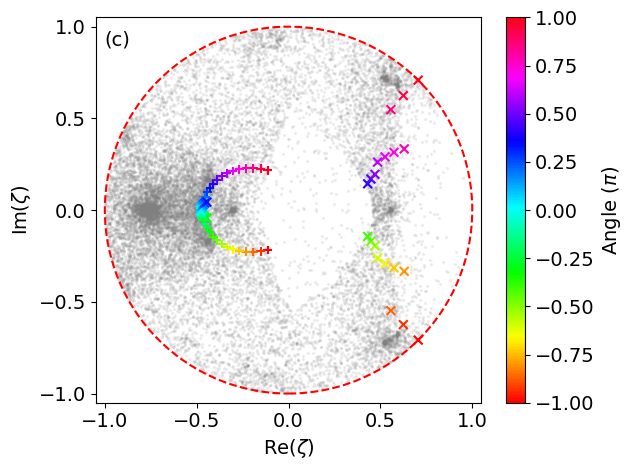}
\includegraphics[width=6cm]{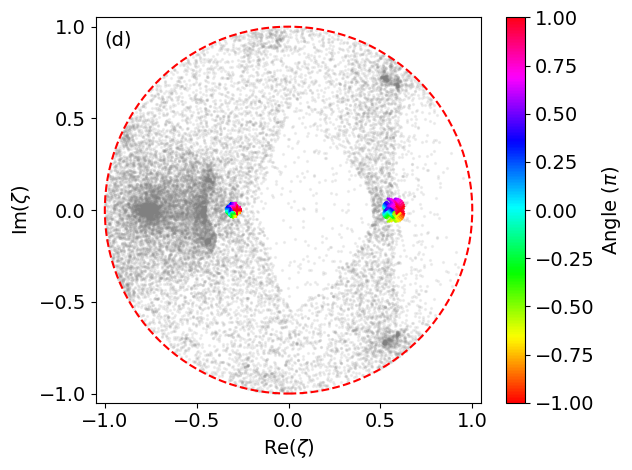}
\end{center}
\caption{
\label{smfig:n3_zeta_evolution}
(a) Evolution of $\zeta$s 
with the asymmetry parameter $\alpha$ when the two electrons are fixed 
along a diameter opposite to each other.
The plus and cross signs correspond to $\zeta$s for the second and third electrons, 
respectively. 
The grey background is the distribution of $\zeta$ for 100,000 random configurations 
of the two fixed electron positions, reproduction of Fig.~\ref{smfig:n3_zero_map}(a).
(b)-(d) Evolution of $\zeta$ with the angle $\theta$ when one of the electrons rotates 
around the origin. 
We fix $\alpha = 0.1$ in (b), $\alpha = 0.8$ in (c), and $\alpha = 0.98$ in (d).
}
\end{figure*}

First, we consider the case where two fixed electrons are aligned along a diameter
opposite to each other.
For simplicity, we assume the diameter is the real axis, so the electron positions are real,
which can be parametrized as 
\begin{equation}
z_e^{(2)} = 1 - \alpha, \quad z_e^{(3)} = - (1 + \alpha),
\end{equation}
where $0 \leq \alpha < 1$ such that $\chi$ is real.  
When $\alpha = 0$ such that the two electrons are located equidistant to the origin,
$\zeta^{(2)}$ (marked by the plus sign) and $\zeta^{(3)}$ (marked by the cross sign) 
are identical to $\xi_0 = -0.77$, 
coinciding with the highest density peak in Fig.~\ref{smfig:n3_zero_map}(b). 
We confirm the result in Fig.~\ref{smfig:n3_zero_map}(c), in which
we show a randomly selected configuration that has $\zeta = -0.77$.
Not surprisingly, the two electrons in such a configuration are located almost 
symmetrically around the origin. 
As $\alpha$ increases, the two $\zeta$s, as shown in Fig.~\ref{smfig:n3_zeta_evolution}(a), 
agree reasonably well and shift along the real axis to $\zeta = -0.5$, 
where an arc with three weak peaks crosses the real axis in the density map
in Fig.~\ref{smfig:n3_zero_map}(b). 
The synchronized evolution breaks down for $\alpha \gtrsim 0.9$, 
when $\zeta^{(3)}$ jumps to the positive side of the real axis. 
This is because the farther non-Pauli zero wanders too far away from the corresponding Pauli zero, 
such that one of the non-Pauli zeros accompanying another Pauli zero 
on the other side becomes closer.
This should be regarded as a failure in identifying the group of triple zeros, 
but it belongs to rare events with $\vert \chi \vert > 20$;
alternatively, this can be thought of as an edge effect. 
As $\alpha$ approaches 1, the abnormal $\zeta^{(3)}$ approaches 0.56,
another weak peak in the density map. 

We next consider the case where one of the electrons rotates around the origin, 
such that the two electron locations can be parametrized as 
\begin{equation}
z_e^{(2)} = 1 - \alpha, \quad z_e^{(3)} = - (1 + \alpha) e^{i \theta},
\end{equation}
where $0 \leq \alpha < 1$ and $-\pi < \theta \leq \pi$.  
Fig.~\ref{smfig:n3_zeta_evolution}(b) shows the evolution of $\zeta$ with the angle $\theta$
with fixed $\alpha = 0.1$.
For $\theta$ close to 0, $\zeta$s fan out from its value at $\theta = 0$, 
contributing to the surrounding of the peak at $\zeta_0$ in the density map. 
For large angles, however, $\zeta$s spread out all over the unit circle. 
As $\alpha$ increases, the $\zeta$s stretch more towards the positive real axis
and spread less. 
Figs.~\ref{smfig:n3_zeta_evolution}(c) and (d) show the same angular evolution of $\zeta$ 
with fixed $\alpha = 0.8$ and 0.98, respectively.
When $\alpha = 0.8$, $\zeta^{(2)}$ for the electron closer to the origin traces 
a C-shape curve on the left half, while $\zeta^{(3)}$ appears on the other half 
except for small $\theta$. 
When $\alpha = 0.98$ or $\chi \approx 100$, $\zeta^{(2)}$ circles around -0.30 
while $\zeta^{(3)}$ around 0.56, contributing to the two weaker peaks in the density map. 

Comparing the three-electron results with those from larger systems, 
we understand that the $\zeta$ peak at $\zeta_0 = -0.77$ here is of the same origin 
as the $\zeta$ peak in larger systems, e.g., at $\zeta = -0.85$ for $N = 6$ in Fig. 2(c) 
in the main text. 
The numerical difference is due to the fact that in large enough systems 
bulk electrons are likely to be surrounded by electrons at all directions, 
so the non-Pauli zeros tend to locate more symmetrically around the electrons.
For $N = 3$, larger fluctuations in the electron/zero distribution lead to 
$\zeta$s deviating more significantly from $-1$. 

\begin{figure*}[b]
\begin{center}
\vspace{0.5cm}
\includegraphics[width=\linewidth]{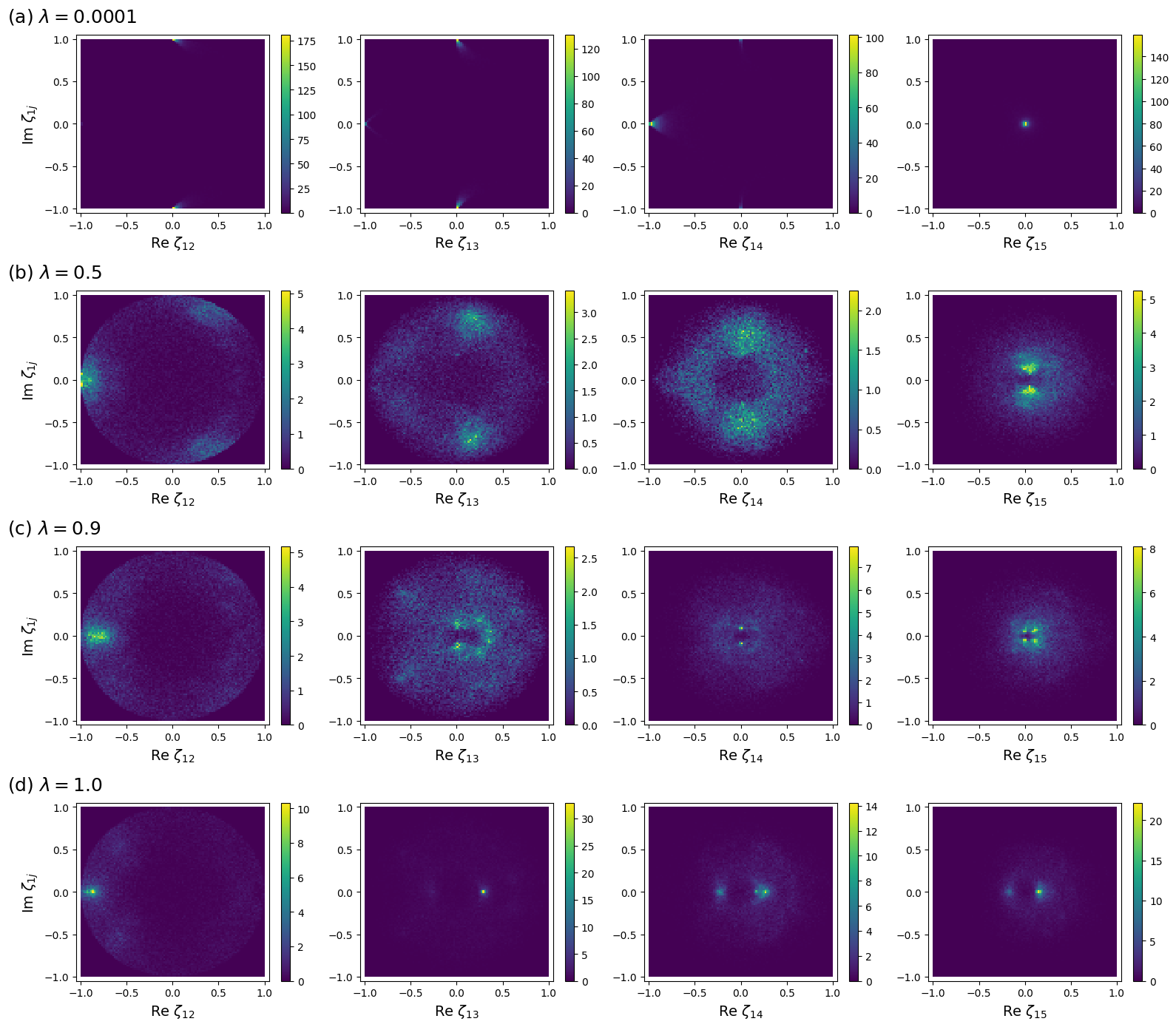}
\vspace{-0.5cm}
\end{center}
\caption{
\label{smfig:one_fifth}
The density of $\zeta_{1j}, j = 2, 3, 4, 5$ for clusters of zeros in $N = 4$ ground states
at $\nu = 1/5$ with mixed $V_1 = V_3 = 1$ and Coulomb interaction $H_C$.
The relative weight $\lambda$ of $H_C$ increases from (a) 0.0001, (b) 0.5, (c) 0.9, to (d) 1. 
}
\end{figure*}

\subsection{II. Evolution of the Zero Displacement Ratio Distribution at $\nu = 1/5$ in Disk Geometry}

For quantum Hall states with more complex zero structures,
we can generalize the definition of zero displacement ratio to
\begin{equation}
    \zeta_{ij} = \frac{z_i - z_e}{z_j - z_e}, \quad i < j. 
\end{equation}
for better understanding more complex zero structures.
Here, $i = 1, 2, \dots$ and $j = 2, 3, \dots$ denote the indices of the $i$th and $j$th 
nearest neighboring ($j$NN) zeros, respectively, for the electron at $z = z_e$.
In the main text, we selected a special example at $\nu = 1/5$ 
in which the four zeros bound to each electron evolve from a star-like structure 
to a linear one as the weight of long-range interaction $\lambda$ increases.
We present the corresponding evolution of $\zeta$ in disk geometry in this appendix. 

At filling fraction $\nu = 1/5$, the Laughlin wave function has 
four zeros bound to each electron. 
They tend to form two equidistant pairs along perpendicular directions,
which are characterized by $\zeta_{12, 13, 14} = \pm i, -1$, 
as shown in Fig.~\ref{smfig:one_fifth}(a) for almost no Coulomb interaction ($\lambda = 0.0001$).
Clearly, each electron binds only four zeros, so $\zeta_{15}$ peaks around 0,
indicating the remoteness of the 5NN zero. 
For $\lambda = 0.5$, the peaks at $\zeta_{12, 13, 14} = \pm i, -1$ broaden and shift away from
these ideal values, as shown in Fig.~\ref{smfig:one_fifth}(b), 
much alike the evolution of the $\nu = 1/3$ case discussed earlier.
Meanwhile, $\zeta_{15}$ broadens and splits towards $\pm i$, as the 5NN zero, 
which belongs to the nearest neighboring electron, 
becomes relatively closer in terms of the shortest electron-zero distance. 
Nevertheless, the main feature of $\zeta_{15}$ remains significantly smaller than those of 
$\zeta_{12, 13, 14}$, indicating that each electron still binds four zeros predominantly.

The situation changes drastically when electrons interact via 
pure Coulomb interaction ($\lambda = 1$),
as shown in Fig.~\ref{smfig:one_fifth}(d). 
The peaks of $\zeta_{12, 13, 14}$ align along the real axis, 
with the corresponding distance of the 3NN and 4NN zeros to the electron 
becoming comparable to that of the 5NN zero.
This suggests that each electron binds only two zeros tightly, 
while the other two appear to meander around farther away.
The four zeros bound to the same electron no longer form 
a square-like pattern around it,
as in the Laughlin-like states at small $\lambda$, 
but rather form a one-dimensional crystal-like structure.
This confirms that the example we gave in the main text 
represents the generic behavior of the system. 

The transition occurs around $\lambda = 0.9$, 
as shown in Fig.~\ref{smfig:one_fifth}(c). 
One unusual feature is that the 3NN zero 
moves away significantly from the electron and 
seems to smear along all directions, 
suggesting that the cluster of four zeros 
breaks apart with only two still tightly bound. 
On the other hand, $\zeta_{14}$ develops two density peaks 
along the imaginary axis not far from the origin, reminiscent of the Laughlin order but at a much larger distance 
than that of the tightly bound ones. 
The residual Laughlin order is also reflected in the four-peak density 
of $\zeta_{15}$ at $\lambda = 0.9$.

\begin{figure*}
\begin{center}
\vspace{0.5cm}
\includegraphics[width=12cm]{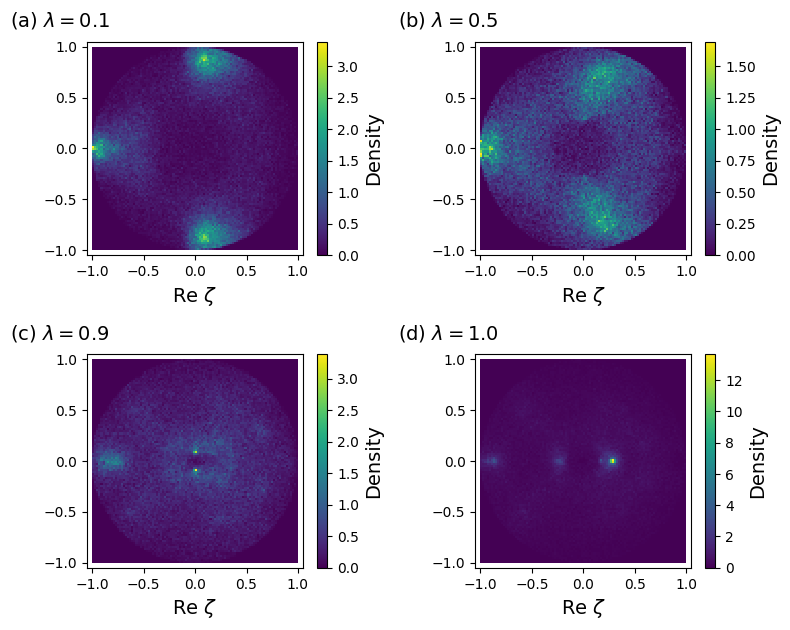}
\vspace{-0.5cm}
\end{center}
\caption{
\label{smfig:one_fifth_sum}
The total density of $\zeta_{12, 13, 14}$ for clusters of zeros in $N = 4$ ground states
at $\nu = 1/5$ with mixed $V_1 = V_3 = 1$ and Coulomb interaction $H_C$.
The relative weight $\lambda$ of $H_C$ increases from (a) 0.0001, (b) 0.5, (c) 0.9, to (d) 1. 
}
\end{figure*}

The evolution and the transition can also be visualized by 
adding the densities of $\zeta_{12, 13, 14}$ together, 
as shown in Fig.~\ref{smfig:one_fifth_sum}.
The evolution pattern, again, strongly suggests the 
usual picture of the Laughlin phase at $\nu = 1/5$ that 
each electron evenly binds four zeros cannot be stabilized 
by Coulomb interaction, whose long-range part tends to arrange the zeros 
into a one-dimensional lattice to maximally reduce their interaction. 
As a result, two zeros or vortices are tightly bound to each electron consistently, 
forming a CF in an effective filling fraction $\nu^* = 1/3$.  
The change in the strength of the Coulomb interaction $\lambda$ 
tunes the range of the residual interaction, which decides 
the arrangement of the remaining zero.
For $\lambda \sim 1$, CFs have a relatively long-range residual interaction, 
the binding of the additional zeros is weak, 
and they tend to align along the core CFs. 
As $\lambda$ decreases, CFs eventually interact via a short-range interaction 
and form a more Lauglin-like liquid, 
i.e., the second pair zeros become closer to the first pair 
along the perpendicular direction. 
The hierarchy of zeros eventually disappears as $\lambda$ approaches 0, 
resulting in the Laughlin state at $\nu = 1/5$. 

\subsection{III. The composite fermion fluid wave function for Coulomb interaction at $\nu = 1/5$ in disk geometry}

For $\nu = 1/5$, the conventional model wave function is the Laughlin wave function, 
whose parent state is the $\nu = 1$ integer quantum Hall (IQH) state, i.e.,
\begin{equation}
\Psi_{\rm Laughlin} \equiv \prod_{i<j} (z_i - z_j)^5 = \prod_{i<j} (z_i - z_j)^4 \Psi_{\rm IQH}(\{z_i\}),
\end{equation}
Each electron is attached with 4 vortices, making them $^4$CFs. 
Naively, with long-range Coulomb interaction, the resulting four zeros 
around each electron smear out symmetrically. 
However, the overlap between the Coulomb ground state and the Laughlin wave function 
deteriorates significantly as the system size increases. 
Even for $N = 6$ electrons, the overlap amplitude is less than 0.85 in disk geometry, 
so the overlap probability 
\begin{equation}
\left \vert \left \langle \Psi_{\rm Laughlin} \vert \Psi_{\rm Coulomb} \right \rangle 
\right \vert^2 \approx 0.7. 
\end{equation}
This is worse than the $\nu = 1/3$ case, 
which, in retrospect, hints that the smearing of the two pairs of zeros may not be symmetric. 

The earlier work of Chang et al.~\cite{chang06} in some sense is pursuing the same goal 
of optimizing the wave function to model a realistic situation. 
The difference is that,
even though the Coulomb ground state is believed to be a quantum liquid, 
they started from an $N$-electron crystallite parent state $\Psi_{\rm EC}^{N, L^*}(\{z_i\})$ in disk geometry
at angular momentum $L^* = 3N(N-1)/2$ at $\nu^* = 1/3$ effective filling. 
The parent state comes from a Hartree-Fock electron crystal, 
in which electrons are localized in LLL coherent-state wave packets, 
and which is projected to the total angular momentum $L^*$~\cite{chang06}.
The lattice can be triangular, or more complicated ones for finite systems. 
Then, each electron is attached with 2 vortices, making them $^2$CFs 
in the CF crystal state
\begin{equation}
    \Psi_{\rm CFC} = \prod_{i<j} (z_i - z_j)^2 \Psi_{\rm EC}^{N, L^*}(\{z_i\}),
\end{equation}
pushing the true filling down to $\nu = 1/5$. 
Interestingly, the overlap of such CF crystal state with the Coulomb ground state 
improves significantly over that of the Laughlin state, 
because the construction properly takes care of the crystalline order, 
which we demonstrated by the density map of the zero displacement ratio
in the previous section. 
The twist, thus, needs a justification that the liquid state prevails 
in the thermodynamic limit. 
For that, Chang et al.~\cite{chang06} showed in their work that for $N \geq 10$ electrons, 
the Laughlin wave function has a lower energy. 

The CF fluid (CFF) state we introduce is in the same spirit, except that the parent state 
is a fluid state. It has the form 
\begin{equation}
    \Psi_{\rm CFF} = \prod_{i<j} (z_i - z_j)^2 \Psi_{\rm fluid}^{1/3}(\{z_i\}),
\end{equation}
where $\Psi_{\rm fluid}^{1/3}$ is of the same symmetry as the targeted $\nu = 1/5$ state.
The parent state can be the ground state of the Coulomb interaction or its variance $(1-\lambda)V_1+\lambda H_C$
at angular momentum $L^* = 3N(N-1)/2$ at $\nu^* = 1/3$. 
The essence of such a construction is not to find a new topological phase, 
but to allow a meaningful tuning of the wave function to accommodate 
the geometrical evolution among a family of wave functions of the same topology. 
As we showed in the main text, the construction allows us to approximate 
the Coulomb ground state with high precision at small system sizes,
in which we afford the exact brute-force expansion of the multiplicative factor 
of the flux attachment as a sum of $3^{N(N-1)/2}$ monomials 
and apply them to each basis state independently.
This also means that we can calculate the overlap with the Coulomb ground state efficiently 
by vector inner product, instead of Monte Carlo sampling that is necessary in larger systems. 

\begin{figure*}
\begin{center}
\includegraphics[width=\linewidth]{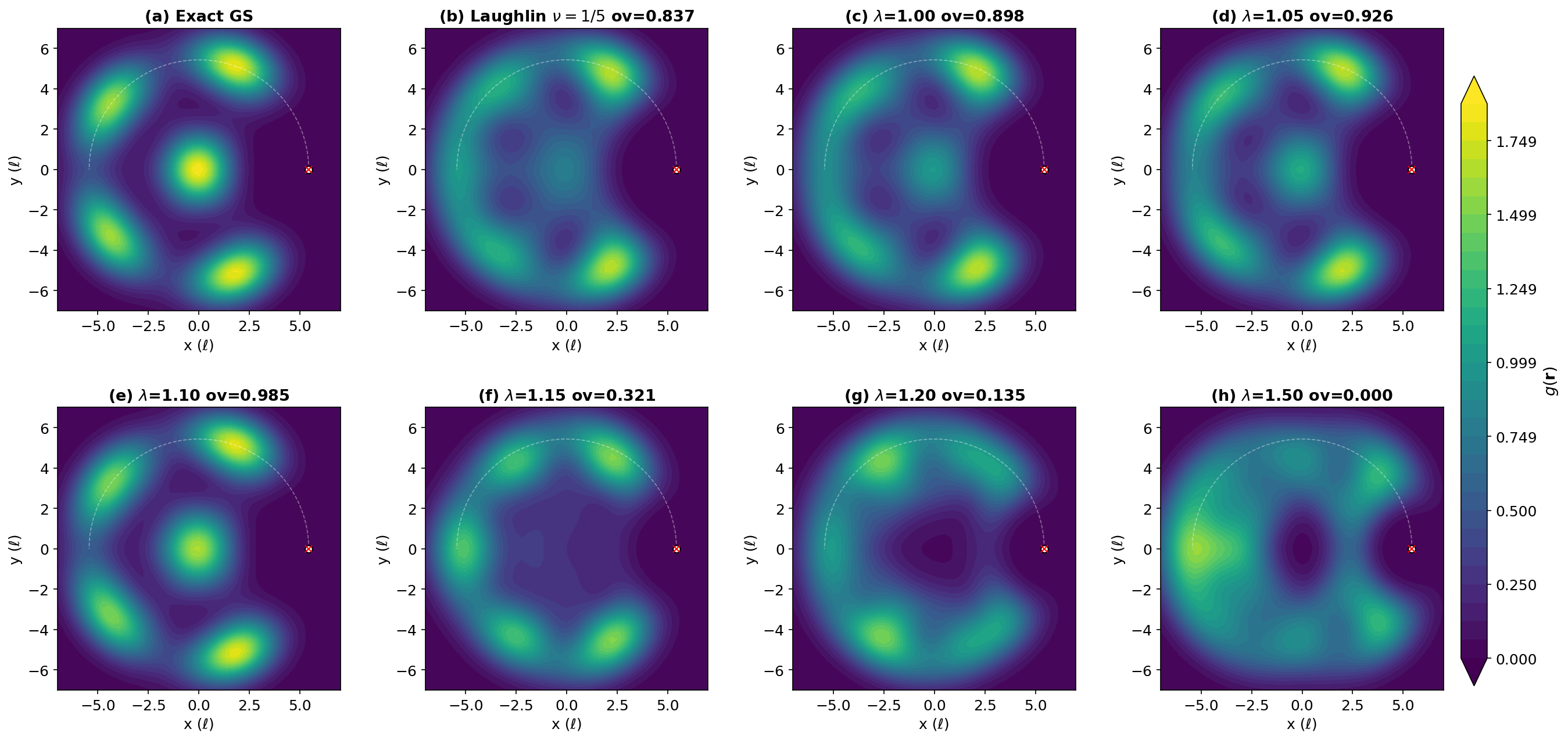}
\end{center}
\caption{
\label{smfig:n6_2d_map}
Pair correlation function for an $N = 6$ system at $\nu = 1/5$. 
The wave functions are (a) the exact ground state, 
(b) the Laughlin wave function, (c)-(h) the CFF wave functions
whose parent states are the ground states of the mixed Coulomb and hardcore interaction,
with mixing parameter $\lambda = 1.0, 1.05, 1.1, 1.15, 1.2, 1.5$.
An electron is fixed at $R = 5.434 \ell$ (indicated by the red cross), 
where the electron density is the highest in the exact ground state.
Here, the exact pair correlation function is closest to that of the CFF wave function 
with $\lambda \approx 1.1$.
}
\end{figure*}

To compare our construction with the earlier work of Chang et al.~\cite{chang06}
and to understand the evolution of the CFF wave functions, 
we calculate the pair correlation function 
\begin{equation}
g(\mathbf{r}, \mathbf{R}) \sim \int \prod_{j = 3}^{N} d\mathbf{r}_j 
\vert \Psi_{\rm CFF}(\mathbf{r}, \mathbf{R}, \mathbf{r}_3, \dots, \mathbf{r}_N) \vert^2
\end{equation}
for the CFF wave functions at $\nu = 1/5$. 
Fig.~\ref{smfig:n6_2d_map} compares $g(\mathbf{r}, \mathbf{R})$ for $N = 6$ CFF wave functions
with $\lambda = 0.0$ (Laughlin), 1.0 (Coulomb), 1.05, 1.1, 1.15, 1.2, 1.5,
where we choose $\mathbf{R} = (5.434 \ell, 0)$ as the electron desity peak of the exact ground state,
which is marked by a red cross in Fig.~\ref{smfig:n6_2d_map}.
The striking similarity between the pair correlation function at $\lambda = 1.1$ 
and that of the exact Coulomb ground state is consistent 
with the high overlap amplitude 
\begin{equation}
\mathcal{O} (\lambda) = \left \vert \left \langle \Psi_{\rm CFF} (\lambda) \vert 
\Psi_{\rm exact} \right \rangle \right \vert 
\approx 0.985,
\end{equation}
a significant increase from 0.837 for $\lambda = 0.0$. 
The overlap increases gradually with $\lambda$ until $\lambda \approx 1.1$.
Meanwhile, the pair correlation function for $\lambda = 1.0$ and $1.05$ 
is similar to that for the Laughlin state, 
confirming their fluid nature. 
Note that when the maximum overlap is reached, 
$g(\mathbf{r}, \mathbf{R})$ develops five lumps, one at the origin and 
the other four along the semicircle, due to the increasing Coulomb repulsion.
However, such an underlying electron lattice configuration
is not explicitly used in our approach, but is needed for the CF crystal construction. 
Interestingly, the pattern ceases to exist when $\lambda \geq 1.15$.
In particular, the pair correlation function almost vanishes near the origin, 
indicating the destruction of the liquid phase that tends to maintain uniform electron density. 
The transition occurs around $\lambda = 1.15$ with $\mathcal{O} = 0.321$,
where $g(\mathbf{r}, \mathbf{R})$ exhibits an underlying ring configuration. 

\begin{figure*}
\begin{center}
\includegraphics[width=\linewidth]{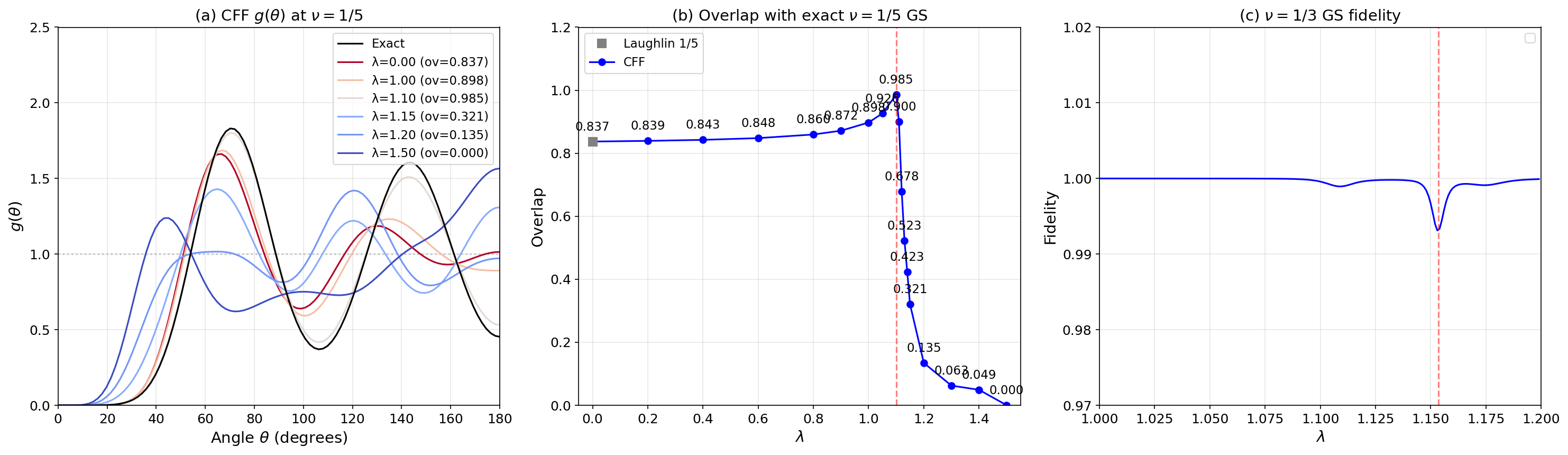}
\end{center}
\caption{
\label{smfig:n6_scan_gtheta}
(a) The pair correlation function for $N = 6$ CFF wave functions at $\nu = 1/5$
whose parent states are the ground states of the mixed Coulomb and hardcore interaction,
with mixing parameter $\lambda = 0.0, 1.0, 1.1, 1.2, 1.5$.
The exact pair correlation function agrees best with the CFF wave function 
with $\lambda \approx 1.1$.
We fix an electron at $R = 5.434 \ell$ and plot the pair correlation function 
as a function of the radial angle $\theta$ along the semicircle of radius $R$ 
(marked by the dashed line in Fig.~\ref{smfig:n6_2d_map}). 
(b) The overlap amplitude between the CFF wave function and the Coulomb ground state 
at $\nu = 1/5$ as a function of $\lambda$. 
The maximum overlap is achieved at $\lambda \approx 1.1$.
(c) The fidelity of the ground state of the mixed Hamiltonian at $\nu^* = 1/3$ 
as a function of $\lambda$. 
The fidelity develops a dip at $\lambda \approx 1.15$, signaling a phase transition 
in the parent state from the topological liquid phase to an ordered phase.
}
\end{figure*}

For further understanding, in Fig.~\ref{smfig:n6_scan_gtheta}(a) we plot the angular 
dependence of the pair correlation function along the semicircle of 
radius $R = 5.434 \ell$, marked by the dashed line in Fig.~\ref{smfig:n6_2d_map}.
One clearly observes a transition from the liquid phase, 
in which the first peak dominates, 
to a charge-density wave phase,
in which the peak height alternates.
The latter density oscillation is clearly observable in Fig.~\ref{smfig:n6_2d_map}(g) and (h).
The dependence of the overlap amplitude with the Coulomb ground state 
is shown in Fig.~\ref{smfig:n6_scan_gtheta}(b) as a function of $\lambda$. 
Again, the maximum overlap is achieved at $\lambda \approx 1.1$. 
Remarkably, $\mathcal{O}$ drops slowly with decreasing $\lambda$ on the left side,
while sharply with increasing $\lambda$ on the right side.
The overlap peak and the striking contrast in the CFF wave functions can be 
traced to a transition in their $\nu^* = 1/3$ parent states.  
To reveal the connection, 
we plot in Fig.~\ref{smfig:n6_scan_gtheta}(c) 
the fidelity of the ground state 
\begin{equation}
    f(\lambda) = |\langle\psi(\lambda)|\psi(\lambda+\Delta\lambda)\rangle|
\end{equation}
of the mixed Hamiltonian at $\nu^* = 1/3$ 
as a function of $\lambda$. 
Here, we choose $\Delta \lambda = 0.001$. 
The fidelity develops a dip at $\lambda \approx 1.15$, signaling a phase transition 
in the parent state from the topological liquid phase to a distinct phase.
The relative weakening of the $V_1$ component for $\lambda > 1$ 
has been first studied by Haldane and Rezayi~\cite{haldane1985}. 
They found that, in spherical geometry the transition occurs at $\lambda_c = 1.25$ and conjectured 
the state after transition is a lattice state. 

The joint evolution of the parent state and the CFF wave function reveals that 
the reason that the Laughlin description falls short is due to the 
geometrical distortion of the internal flux-attachment structure 
of the composite fermions induced by the long-range Coulomb interaction.
Such a distortion, when large enough, 
leads to the destruction of the topological liquid phase, 
and hence the transition to the corresponding ordered phase.
By gradually increasing $\lambda$
we observe such an evolution in the parent state,
and the accompanying improvement of the CFF wave function in the approximation 
of the Coulomb ground state, as we approach the transition point. 
The improvement comes from the missing charge-density wave component 
in the neighboring phase
that complement the liquid phase to recover the characteristic strong oscillations 
in the pair correlation function in the presence of the Coulomb interaction. 
The $^2$CF flux attachment shifts the optimal $\lambda$ slightly 
below the transition point $\lambda_c$ in the parent state, 
such that the corresponding parent state is still a liquid, 
but with significant charge-density oscillations due to its proximity to the transition point. 
Such a mixture of neighboring phases has already been demonstrated in the geometrical evolution of 
FQH wave function by principal component analysis~\cite{jiang20},
a dimension reduction tool widely used in machine learning.

\end{document}